**Tradition and Innovation in Scientists' Research Strategies**

Jacob G. Foster[1], Andrey Rzhetsky[2,3], James A. Evans[1,2]

[1]Department of Sociology, University of Chicago, 1126 East 59th Street, Chicago, IL 60637

[2]Computation Institute, University of Chicago, Searle Chemistry Laboratory, 5735 South Ellis Avenue, Chicago, IL 60637

[3]Departments of Medicine and Human Genetics, Institute for Genomic and Systems Biology, 900 East 57th Street, Chicago, IL 60637
University of Chicago

**Abstract**

What factors affect a scientist's choice of research problem? Qualitative research in the history, philosophy, and sociology of science suggests that this choice is shaped by an "essential tension" between the professional demand for productivity and a conflicting drive toward risky innovation. We examine this tension empirically in the context of biomedical chemistry. We use complex networks to represent the evolving state of scientific knowledge, as expressed in publications. We then define *research strategies* relative to these networks. Scientists can introduce novel chemicals or chemical relationships—or delve deeper into known ones. They can consolidate existing knowledge clusters, or bridge distant ones. Analyzing such choices in aggregate, we find that the distribution of strategies remains remarkably stable, even as chemical knowledge grows dramatically. High-risk strategies, which explore new chemical relationships, are less prevalent in the literature, reflecting a growing focus on established knowledge at the expense of new opportunities. Research following a risky strategy is more likely to be ignored but also more likely to achieve high impact and recognition. While the outcome of a risky strategy has a higher expected reward than the outcome of a conservative strategy, the additional reward is insufficient to compensate for the additional risk. By studying the winners of 137 different prizes in biomedicine and chemistry, we show that the occasional "gamble" for extraordinary impact is the most plausible explanation for observed levels of risk-taking. Our empirical demonstration and unpacking of the "essential tension" suggests policy interventions that may foster more innovative research.

**Introduction**

How do scientists choose a research problem? Many factors influence the decision, from past interests and training (1) to serendipitous encounters with salient expertise and information (2). Conflicting professional pressures intensify this choice. Career advancement requires reliable productivity, typically achieved by incremental contributions to established research directions. High standing in the scientific community, by contrast, requires important, original contributions, often achieved by pursuing risky new directions (3). These conflicting demands create a tension between two broad strategies: productive "tradition" and risky "innovation" (4). When following the conservative strategy and adhering to the research tradition in their domain, scientists achieve publication with high probability, remaining visibly productive but forgoing opportunities for originality. When following a risk-taking strategy, scientists fail more frequently, and may appear unproductive for longer periods (3)[1]. If a risky project succeeds, however, it may have profound impact, generating substantial new knowledge and winning broad acclaim (5). This strategic tension mirrors other well-known dichotomies: reliable "succession" versus risky "subversion" (6) or "relevance" versus "originality" (7) in the sociology of science; "conformity" versus "dissent" or "discipline" versus "rebellion" in the philosophy of science (8); and "exploitation" versus "exploration" or "refinement" versus "risk taking" in the study of innovation (9). Recent theoretical work supports this picture of an "essential tension" (4), by highlighting the distinctive contributions (10) and rewards (11) associated with conservative and risk-taking strategies.

Such scholarship suggests that *more important* findings tend to be *more unexpected*, given available knowledge and established methods of inquiry. Established methods tend to produce few surprises, and encourage neglect of unexpected findings when they occur. By contrast, Thomas Kuhn's *Structure of Scientific Revolutions* illustrated the key role of unexpected findings in supplanting scientific theories (5). But how prevalent are the strategies that yield unexpected findings? And what risk and reward are associated with them? Qualitative and theoretical research suggests that such strategies should be rare, risky, and highly rewarded. In this article, we examine these questions quantitatively, in the context of biomedical chemistry. First, we define and measure the prevalence of five broad *strategies* that scientists use to select chemical relationships for study. We perform this analysis in aggregate, pooling the articles published each year and counting strategies observed in each pool. We demonstrate that the relative frequency of these strategies is unrelated to available opportunities to pursue them. Indeed, we find increasing focus on established knowledge within biomedicine over time. Then we link strategies to the average scientific importance of the findings they produce, and discover that rare strategies tend to produce more important findings. Finally, we consider how distinct reward mechanisms influence the selection of scientific strategies. We show that ordinary rewards through citations are insufficient to compensate scientists for their risky gambles. We then analyze the distribution of strategies in articles associated with extraordinary achievement (measured through very high citation counts and prizes), and suggest that the potential for such achievement may motivate scientists to observed levels of risk-taking.

---

[1] Examples include Andrew Wiles while proving Fermat's Theorem or Frederick Sanger while developing the "Sanger method" for DNA sequencing.

Most quantitative analyses of large-scale scientific behavior use the document as the fundamental unit of analysis (12, 13) and leverage citation data to identify innovations (14) and outline the structure of scientific fields (15). We adopt a content-based approach that uses basic conceptual entities (in this case, chemicals) and their relationships as the fundamental units of analysis. These relationships weave a network of research possibilities, corresponding to Simon and Newell's notion of "the space of search paths" or "network of possible wanderings" involved in human cognition ((16), p.82). Under this metaphor, scientists move through the network of knowledge and add to it through their research efforts.

We propose the following coarse taxonomy of research strategies, corresponding to structurally distinct contributions to the network of scientific knowledge. Figure 1 illustrates how a scientist facing a network of known scientific relationships could choose to make a *jump*, *new consolidation*, *new bridge*, *repeat consolidation*, or *repeat bridge* (17, 18). A researcher may propose a relationship involving completely unexplored entities, making a *jump* beyond current knowledge (17)[2]. She may also test a relationship between previously explored entities, either providing a *new* (not previously published) relationship or *repeating* a relationship proposed before. *New* and *repeat* relations, in turn, can be either *consolidations* or *bridges*. This distinction rests on the fact that the knowledge network partitions naturally into clusters of closely related chemicals, mirroring scientific subfields (see *Methods* for an explanation of how we identify these clusters). Joining entities within the same cluster provides further *consolidation* of the cluster, deepening knowledge in that particular domain. Linking entities from distinct knowledge clusters creates a *bridge* between them, altering the broad connectivity of the knowledge network and weaving loosely connected regions of knowledge together more tightly.

**Results and Discussion**

Our network is built from 6,455,756 abstracts in the National Library of Medicine's MEDLINE collection, published 1934 – 2008 and annotated by NLM with two or more chemical entities; see *Materials and Methods*. Chemical annotations were introduced in 1980 and applied to all journals indexed subsequently, accounting for the pre-1980 annotations. We focus analysis on 1983 – 2008 to limit the effect of the introduction of indexing on our results; see *SI*. We construct an evolving network (15, 18) using these annotated articles. By 2008, the network has 181,078 nodes and 84,709,977 links. Nodes represent chemicals; if two chemicals appear in the same article in year *t*, we infer a relationship that connects the chemicals and add a link between the corresponding nodes in the network for year $t+1$ (19). For example, a 2007 article (20) includes a *new consolidation* within a neurobiology-related subfield, showing that neuronal tubulin-preferring agent NAPVSIPQ, a neuroprotective oligopeptide, defends neurons against neurotoxic kainic acid. Network links between chemicals persist (although this assumption can be relaxed; see *SI*), and chemicals can be linked repeatedly. Note that articles are often annotated with more than two chemicals, and therefore contribute more than one chemical relationship. All relationships appearing in a given year enter the network together, as if published simultaneously. While an approximation, this keeps the network evolution tractable. This procedure yields a time-ordered sequence of weighted networks over chemicals. In a given year, we characterize the links associated with each publication using one of the five strategies. This characterization allows us to count the total number of times we observe each strategy in a given year.

---

[2] She may explicitly set out to design a new compound (e.g., creating a new pharmaceutical) or may find evidence of a new entity and set out to characterize it (e.g., describing a new receptor or antibiotic).

**Prevalence and stability.** Pooling these counts across all years, we find that the frequency of each strategy in the published literature from 1983 – 2008 appears to be inversely related to its plausible risk of failure. *Repeat* statements were six times more frequent than *new* or *jump* statements (85.8% versus 14.2%). *New bridges* and *new consolidations* were more common than *jumps* by roughly the same proportion (12.4% versus 1.8%). It is likely that strategies that build incrementally on former knowledge appear more frequently not only because they are pursued more frequently, but also because non-incremental strategies are prone to fail, remaining unpublished and invisible to science; see *SI*[3].

Using the strategy counts for a given year, we can also compute their relative frequency within that year. The resulting distribution of scientific strategies has remained remarkably stable over the period studied; see Figure 2A (solid lines) and *SI*. From 1983 – 2008, the fraction of total links in a given year corresponding to "new" strategies has changed little, with *jumps* the rarest, followed by *consolidations*, then *bridges*.

This stability is surprising. Biomedical chemistry has changed enormously over the past few decades. For example, the number of distinct chemicals that appear in MEDLINE annotations increased by an order of magnitude between 1980 and 2008 (see Fig S1). Because of this growth, there are vastly more new links to explore than known links (a factor of 22 more in 1983, growing to a factor of 188 by 2008). Likewise, because knowledge clusters are small relative to the entire system (the largest shrinks from 27.5% of the network in 1983 to 12.8% in 2008) the number of possible bridging links grows faster than the number of possible consolidating links (see Fig. S4). The pattern and stability of strategic choices suggests that scientists are not responding to the changing space of research possibility. Even if an overwhelming fraction of potential new links can be discarded reliably as physically impossible or otherwise unworthy of exploration— without doing a single experiment—scientists are still not taking full advantage of opportunities to explore new relationships between chemicals.

---

[3] Non-incremental strategies, when successful, may be easier to publish in high-prestige outlets (as their results are considered more novel). Nevertheless, we believe the results of incremental research are relatively easy to publish, albeit in lower-impact journals.

**Stability and attention.** The stability of the strategy distribution despite dramatic changes in research opportunities suggests that scientific attention may be narrowing in focus. To explore this, we introduce a simple, generative model of strategy selection (17). In this model, scientists (represented by a single abstract agent) independently and randomly choose a strategy at time $t$ based on the number of available links corresponding to each strategy at time $t-1$. The choices of this agent are weighted using *bias parameters* that coarsely summarize how much more salient some opportunities are than others: $\nu$ and $\rho$ are the biases for new and repeat relationships, $\theta_n$ and $\theta_r$ are biases for new and repeat consolidations, and the bias for jumps is normalized to 1 (*Materials and Methods*). We assume that our agent first chooses between *jump*, *new*, or *repeat* links—the most risk-relevant distinctions—and then makes a secondary decision between contributing a *consolidation* or *bridge*. Other model formulations give similar results.

Conceptually, we imagine that the agent repeats this procedure until it has made the same number of choices we observe in a given year. Using the aggregate history of strategic choices discussed above (i.e., the number of times each strategy appears in a given year), we infer bias parameters that maximize the likelihood our agent will generate the observed history. The bias parameters characterizing the agent thus provide an "average" description of the preferences of the scientific community for the various strategic opportunities.

Figure 2A (dashed line) shows that this model predicts observed behavior reasonably well, with a high correlation between known and predicted values (Pearson's $R = 0.983$). The significant trends in parameters suggest that scientists filter out more new opportunities and become more locally focused as knowledge grows. Figure 2B (left axis) indicates a sharply increasing preference for repeating known links instead of exploring new ones. Figure 2B (right axis) shows that the preference for new consolidation over bridging has grown, and may be leveling off, indicating a similar local focus in the exploration of new relationships.

**Risk and rewards.** We turn next to our claim that rare, risky strategies should be more highly rewarded. Figure 3 shows that rare and risky strategies, like *jumps* and new *consolidations*, are indeed rewarded with more citations. In information theory, the self-information or *surprisal* $I(p_i) = \log(1/p_i) = -\log(p_i)$ measures the information associated with observing outcome $i$ of a discrete random variable (21). Highly improbable outcomes are surprising, so this measure assigns a large value to low probability outcomes and quantifies how "surprising" they are. In our case, the outcomes are observations of a given research strategy; observations of rare strategies are more surprising. For a given year, we can define the surprisal of each strategy using our generative model. If $P(strategy)$ is the probability that the representative agent will choose a given strategy in a given year, then the surprisal of that strategy is $-\log P(strategy)$). Strategy surprisal positively predicts the mean citations received by papers containing at least one instance of that strategy during the period 1983-2002, for which we have reliable citation data[4]. When a strategy is an order of magnitude less probable (e.g., 0.5 vs. 0.05), it will receive 2.26 additional citations, on average. In fact, surprisal explains 29.2% of the variation in mean citations (*Materials and Methods*). The size of this effect supports our proposal that important findings tend to result from surprising strategies. Indeed, the relationship is robust to a variety of tests, including distinct subfield definitions, use of median rather than mean citations, and tests that explicitly account for the overdispersion of citation data; see *SI*. We also confirm the intuition that rare strategies tend to be riskier, in the sense that their citation impact is less predictable: the standard deviation in citations is higher for rare strategies. Strategy surprisal explains 28.6% of variation in the standard deviation in citations to papers containing at least one instance of that strategy. This result implies that rare, riskier strategies are often less successful at garnering attention than more conservative strategies. When they do attract attention, however, they attract more attention (i.e., citations) than conservative strategies.

But do rare, risky strategies attract enough attention to balance the chance of utter failure (i.e., no publication)? In other words, is a scientist who chooses a risky strategy being rational? We cannot directly observe from our data the probability that a particular strategy will succeed or fail; the links in our sample come from projects that *did* produce a publication. We can provide bounds on the failure probability, however, if we assume that scientists select strategies only to maximize citations (6). To perform this risk analysis, we extend our strategic taxonomy to "project" strategies characterizing entire articles[5]. Scientists may pursue a *repeat* project with no new relationships or chemicals; a *new* project with at least one *new* relationship but no new chemicals; or a *jump* project with at least one new chemical. The same relationship that links rare strategies with citations also holds for these project strategies (see *SI* for a detailed analysis.) The expected citation count for a particular strategy $E(S)$ is the product of the probability of success $p(S)$ with the mean citations for articles employing that strategy. Our use of mean citations (22) reflects the simplest possible utility function for scientists with no risk-aversion or risk-seeking. We compute mean citations from the data, averaging over the period $1983 - 2002$: 8.38 for *repeat,* 11.00 for *new*, and 12.90 for *jump*. For a risk-neutral scientist to be indifferent between strategies, i.e., for the choice of *new* or *jump* instead of *repeat* to be rational, then $E(repeat) = E(new) = E(jump)$. This equality implies that $p(new) = 0.76 \cdot p(repeat)$ and $p(jump) = 0.65 \cdot p(repeat)$. *A priori*, it seems unlikely that a *new* or a *jump* project is only 24% or 35% less likely to succeed than a conservative *repeat* project. This argument suggests that *repeat* is very likely to be the dominant strategy from a rational choice perspective, i.e., the one that maximizes expected citations. This confirms one "half" of the essential tension. Scientists are strongly incented by systems of professional evaluation, which often rest on citations, to pursue conservative *repeat* strategies. But we are left with a puzzle: while scientists eschew *most* opportunities for high-risk, high-impact work, they still engage more risk than if they only sought to maximize citations—for example, if they were only concerned with job security and a stable career.

---

[4] We can actually estimate the *optimal* fraction of links in a paper corresponding to a given strategy, i.e., the fraction that maximizes citations; see *Supporting Information* on strategy mixtures.

[5] To perform this analysis, we need to associate each article to a unique strategy, hence the "project"-level strategies.

**Risk and awards.** Beyond job security, scientists are motivated by the desire for *significant* impact and recognition. They wish to leave their mark on scientific history (23). This level of achievement is captured by the most highly cited papers (22); even more so by awards and prizes. This level of recognition is thought to require riskier, original contributions (4, 24). Hence we predict that top cited articles and prize winners will deploy rare, risky strategies more frequently than the typical article or scientist. Figure 4 verifies with an aggregate analysis that top cited articles deploy significantly more novel strategies than all articles (see Figure 4B) as a fraction of links contributed, with the strongest enrichment in jumping and new consolidation. Articles written by authors who have won one of 137 different prizes in biomedicine and chemistry show a similar pattern of enrichment. Articles written by winners of the most elite awards (including Nobel prizes) not only introduce new chemicals more frequently than those written by typical scientists, but also more frequently introduce new relationships *within* knowledge clusters (*Materials and Methods*). A new, integrating linkage within a chemical cluster is likely to attract the attention of the established community whose articles inscribe the connections that define the cluster. Indeed, elite award winners may define or transform our understanding of these clusters. The same pattern obtains when we analyze awards by field, grouping them into biomedicine (Figure 4C) and chemistry (Figure 4D). See Table S2 for a complete list of awards analyzed. These results confirm the other half of the essential tension, linking risky innovation to extraordinary scientific achievement.

## Conclusions

Taken together, our results provide strong quantitative evidence that scientists' choice of research problems is indeed shaped by an "essential tension" between conservative productivity and risky innovation (4). Understanding the research process remains a central challenge for science studies, and improved understanding will be key to improved science policy (25). Science is a complex system (26), and new methods like ours can identify processes that govern its evolution. Despite our novel methods, this study has several limitations. Protocols for chemical annotation at the National Library of Medicine are not uniform over time. Our method necessarily misses relationships whose chemicals are not in the NLM annotation. By considering all co-mentioned chemicals, we allow some "false relationships," failing to distinguish between substantial and incidental associations; see *SI*. Moreover, our taxonomy of research strategies does not reflect the variety of factors scientists consider. Finally, strategies that failed to yield publishable findings are excluded from our analysis by our focus on publications. More precise extraction of chemical relationships (25, 27), a broader full-text corpus including research proposals, and richer models can transcend these limits.

Our research suggests one reason why unexpected findings that change the landscape of science are so infrequent: pursuing them is a gamble, without enough citation payoff, on average, to justify the risk. Nevertheless, science benefits when individuals overcome the strategic preferences that orient them toward established islands of knowledge (17) in the expanding ocean of possible topics. Early breakthroughs in literature-based discovery illustrated the power of linking islands of knowledge together (28). Our results explain why such discovery methods remain so fruitful for science. Our findings also suggest new research questions in the science of science, e.g., how strategies change over a scientific career and whether high-achieving scientists initiate more risky projects, are more far-sighted, or simply luckier.

To be sure, not all scientists should pursue risky strategies. "Normal" science that characterizes a known relationship more deeply has its own value. Nevertheless, stimulating innovation is an important goal of science policy, and we close by suggesting two policy levers to promote risk-taking. Since career pressure is a powerful incentive for conservative behavior, decoupling job security from productivity early on can encourage originality, as was the case at Bell Labs (29). So can funding scientists rather than projects, as at the Howard Hughes Medical Institute[6]. The other lever sits outside the tension between productivity and posterity: funding agencies can lower the barriers to risky projects by funding them more aggressively. These interventions may be able to counter the stable conservatism we uncover here.

**Materials and Methods**

***Knowledge Clusters.*** We identify knowledge clusters using the map equation community detection algorithm. We apply it to the relevant network for each year and select the best result out of 50 randomly seeded iterations (30, 31). This algorithm minimizes the description length of a random walk on the network given a two-level labeling scheme. Heuristically, it partitions the network such that the walker tends to spend a long time within a community before transitioning to another, thereby picking out subsets of nodes with dense intraconnection. It is one of the few algorithms that provides good speed and performance on networks of our size (32). Our knowledge clusters or "subfields" reflect the structure of chemical knowledge as expressed in journal articles: chemicals that appear together frequently in articles will be clustered together. We confirmed the robustness of key results against alternate subfield definitions, including those induced by journal classifications and externally curated ontologies; see *SI*.

***Multinomial Confidence Intervals.*** The data in Figs. 2A, 4B, 4C, and 4D measure the fraction of links corresponding to each strategy in various slices of the network (by year; by citation; by prize winner status). We treat the fractions as estimating the parameters of a multinomial distribution with 5 types and model the data as draws from this distribution. We use the confidence interval proposed in (33) with the tighter bound introduced by Goodman (34). If we observe $n_i$ instances of strategy $i$ out of $N$ observations, the $(1-\alpha)$-confidence interval for the proportion of links using strategy $i$, $p_i$ is given by:

---

[6] These interventions raise further questions about how to identify the promising scientists who receive such support; this is a fascinating research topic in its own right.

$$\frac{\chi^2 + 2n_i \pm \sqrt{\chi^2[\chi^2 + 4n_i(N - n_i)/N]}}{2(N + \chi^2)}$$, where $\chi^2$ is the upper $(\alpha/5) \times 100$-th percentile of

the $\chi^2$-distribution with one degree of freedom. Generically, for a multinomial with $k$ types, this would be the $(\alpha/k) \times 100$-th percentile.

***Generative Model of Attention and Strategy Selection.*** We assume that scientists choose from five possible strategies: *jump*; *new consolidation*; *new bridge*; *repeat consolidation*; and *repeat bridge*. For the purposes of this model, we treat each instance of strategic choice as an independent draw (in reality, these choices are correlated at the article-level, but this complicates the analysis). Thus

$$P(jump) + P(new, consolidation) + P(new, bridge) + P(repeat, consolidation) + P(repeat, bridge) = 1$$

We assume that researchers first choose over *jump*, *new*, or *repeat* and then choose between *consolidation* or *bridge*[7]. Hence the probabilities factorize:

$$P(j) + P(n) \cdot P(c \mid n) + P(n) \cdot P(b \mid n) + P(r) \cdot P(c \mid r) + P(r) \cdot P(b \mid r) =$$
$$P(j) + P(n) \cdot \big(P(c \mid n) + P(b \mid n)\big) + P(r) \cdot \big(P(c \mid r) + P(b \mid r)\big) = 1$$

Once *new* or *repeat* is established, scientists choose either a *consolidation* or a *bridge*; $\big(P(c \mid n) + P(b \mid n)\big) = \big(P(c \mid r) + P(b \mid r)\big) = 1$, and $P(j) + P(n) + P(r) = 1$. These equations fully determine the space of possible events.

Researchers (as modeled by a representative agent) independently and randomly choose a strategy at time $t$ based on the number of possible links corresponding to each strategy in the network at time $t$. This number is weighted according to a *bias parameter* that coarsely summarizes the factors influencing this decision. Let $J_t$ be the number of potential jumps, $C_t$ the number of new consolidations, $B_t$ the number of new bridges, $\hat{C}_t$ the number of repeat consolidations, $\hat{B}_t$ the number of repeat bridges, $R_t = \hat{C}_t + \hat{B}_t$ the number of repeat links, and $N_t = C_t + B_t$ the number of potential new links. Repeat variables track the *total* number of repeat links; if a link has been repeated 3,000 times, it contributes 3,000 to the count. This accounting is consistent with our probabilistic framework; the likelihood that a researcher encounters the opportunity for a repeated link is in direct proportion to its number of repetitions. The probability of each strategy is:

---

[7] Several alternative models can be formulated, including a model that associates a parameter to each strategy or one that makes the choice of jump, consolidation, or bridge primary and new or repeat secondary. While all models fit the data reasonably well, the model presented here gives a good fit while also being easy to interpret and presenting a plausible sequence of decisions.

$$P(j) = \frac{J_t}{\nu N_t + \rho R_t + J_t}$$

$$P(n,c) = P(n) \cdot P(c \mid n) = \frac{\nu N_t}{\nu N_t + \rho R_t + J_t} \cdot \frac{\theta_n C_t}{\theta_n C_t + B_t}$$

$$P(n,b) = P(n) \cdot P(b \mid n) = \frac{\nu N_t}{\nu N_t + \rho R_t + J_t} \cdot \frac{B_t}{\theta_n C_t + B_t}$$

$$P(r,c) = P(r) \cdot P(c \mid r) = \frac{\rho R_t}{\nu N_t + \rho R_t + J_t} \cdot \frac{\theta_r \hat{C}_t}{\theta_r \hat{C}_t + \hat{B}_t}$$

$$P(r,b) = P(r) \cdot P(b \mid r) = \frac{\rho R_t}{\nu N_t + \rho R_t + J_t} \cdot \frac{\hat{B}_t}{\theta_r \hat{C}_t + \hat{B}_t}$$

where $\nu$ is the bias for new relationships, $\rho$ the bias for repeats, $\theta_n$ for new consolidation, and $\theta_r$ for repeat consolidation. The bias for jumps is set to 1; as the number of possible jumps decreases as a function of time, $\nu$ and $\rho$ decrease as a function of time as well. We assume the total number of chemicals to be discovered is 250,000. Experiments with subsets of the data indicate that changing the total number of chemicals changes the numerical value of parameters but not their qualitative behavior.

Parameter values are obtained by maximum likelihood estimation, using the observed number of links of each type in the six previous years to construct the likelihood function. Define $n_C(t)$ as the number of new consolidations *observed* in year $t$, and the others in like manner; then $data[t] = [n_J(t), n_C(t), n_B(t), n_{\hat{C}}(t), n_{\hat{B}}(t)]$. The likelihood of a single year of publication data is:

$$L(data[t]) = P(j)^{n_J(t)} P(n,c)^{n_C(t)} P(n,b)^{n_B(t)} P(r,c)^{n_{\hat{C}}(t)} P(r,b)^{n_{\hat{B}}(t)}$$

and the likelihood of all publication data observed between $t_0$ and $t_f$ is:

$$L(data[t_0, t_f] \mid \nu, \rho, \theta_n, \theta_r) = \prod_{t=t_0}^{t=t_f} L(data[t]) = \prod_{t=t_0}^{t=t_f} P(j)^{n_J(t)} P(n,c)^{n_C(t)} P(n,b)^{n_B(t)} P(r,c)^{n_{\hat{C}}(t)} P(r,b)^{n_{\hat{B}}(t)}$$

where we explicitly show the dependence of the likelihood function on the parameters (Note that we assume a uniform prior on the parameters). The parameters are then given by:

$$\{\nu, \rho, \theta_n, \theta_r\} = \arg\max\{L(data[t_0, t_f] \mid \nu, \rho, \theta_n, \theta_r)\} = \arg\max\{\log\left(L(data[t_0, t_f] \mid \nu, \rho, \theta_n, \theta_r)\right)\}$$

The maximum of the likelihood function is calculated using the downhill simplex algorithm implemented in scipy.optimize, i.e., fmin.

***Credible Intervals.*** We assign Bayesian credible intervals to the parameter estimates in Fig. 2B by Monte Carlo sampling (35). For a given set of parameters $\Theta_i = \{\nu, \rho, \theta_n, \theta_r\}$ and likelihood function $L(data, \Theta_i)$ we explore the likelihood function using the Metropolis-Hastings procedure with sequential parameter updates. We begin at the optimal parameter values calculated by the downhill simplex algorithm. With each iteration, we adjust one parameter, cycling through the parameters in a fixed order. The new parameter is chosen by sampling randomly and uniformly from an interval centered on its current value. Given the current parameters $\Theta_{current}$ and the proposed parameters $\Theta_{proposed}$, we calculate $L(data, \Theta_{current})$ and $L(data, \Theta_{proposed})$. If $L(data, \Theta_{proposed}) > L(data, \Theta_{current})$ we accept the transition; if $L(data, \Theta_{current}) > L(data, \Theta_{proposed})$, we accept the transition with probability $p_{accept} = \dfrac{L(data, \Theta_{proposed})}{L(data, \Theta_{current})}$. The intervals for parameter updates are chosen such that the rejection rate is $\approx 30\%$. At each iteration, we record the current parameter values. We performed 200,000 iterations for each parameter estimate. 95% credible intervals for the parameters are the values containing 95% of the sample. To assign intervals to the ratio in Fig. 2B, we conservatively make the intervals as large as possible given intervals $[\rho_L, \rho_U]$ and $[\nu_L, \nu_U]$ by using $\left[\dfrac{\rho_L}{\nu_U}, \dfrac{\rho_U}{\nu_L}\right]$.

***Regressions of Citation on Surprisal.*** Surprisals for each strategy in a given year ( $= -\log[P(strategy)]$ ) are computed using the probabilities given by our generative model. Citations are assigned to articles by linking the MEDLINE abstracts to the ThomsonReuters citation database; each article is then assigned the total citations it received in the three years after its initial publication (the typical half-life for most citation data). If the article is not in the ThomsonReuters database, it is omitted from this analysis. Approximately 2/3 of the MEDLINE articles link to a ThomsonReuters record with citation information. We restrict analysis to 1983-2002 due to decreased citation coverage post 2005.

We use the surprisal for a strategy in a given year to predict the mean citations and standard deviation in citations received by articles in that year with at least one instance of that strategy. The data display some heteroscedasticity, with more scatter in the residuals for larger values of surprisal. To correct for this, we perform regressions in STATA using the HC3 heteroscedasticity correction. In all regression models, the two-tailed *p*-value for the null hypothesis (i.e., a regression coefficient is 0) is less than 0.0001. Full tables of coefficients, t-statistics, and confidence intervals are given in the *SI*.

Surprisal is a highly significant predictor of mean citations, $F(1,98) = 30.96$; $(\text{Prob} > F) < 0.001$; $R^2 = 0.2916$, where $(\text{Prob} > F)$ is the *p*-value of the observed F statistic[8]. The majority of remaining variation is explained by the average increase in mean citations as a function of time; see Figure S7A and B. Year alone is a significant predictor of mean citations: $F(1,98) = 143.62$; $(\text{Prob} > F) < 0.001$; $R^2 = 0.5212$. A combined model including both surprisal and year as predictors captures 76% of the variation in mean citations, F(2,97) = 103.36; (Prob > F) < 0.001; $R^2 = 0.7602$, with surprisal remaining a highly significant predictor, F(1,97) = 51.87, (Prob > F) < 0.001.

---

[8] We find the same behavior if we drop jumps, which have the highest surprisal, from the regression; see *SI* for this and many other robustness checks.

Our findings are similar for standard deviation in citations. Surprisal is a highly significant predictor: F(1,98) = 24.68; (Prob > F) < 0.001; $R^2 = 0.2862$. Year is a significant but weaker predictor: F(1,98) = 33.57; (Prob > F) < 0.001; $R^2 = 0.1895$. A combined model captures 44.4% of the variation, F(2,97) = 26.62; (Prob > F) < 0.001; $R^2 = 0.4442$; surprisal remains highly significant, F(1,97) = 25.36, (Prob > F) < 0.001.

***Prize Winners.*** We compiled a list of 137 different prizes and awards in biomedicine and chemistry, drawing on the category pages for biology awards, medicine awards, and chemistry awards on Wikipedia:

http://en.wikipedia.org/wiki/Category:Biology_awards;
http://en.wikipedia.org/wiki/Category:Medicine_awards
http://en.wikipedia.org/wiki/Category:Chemistry_awards

The majority of prestigious (and many less prestigious) prizes are listed, providing a broad sample of achievement—from field-specific (e.g., the Anselme Payen award, related to cellulose chemistry) to world-recognized (e.g., the Nobel Prize in Physiology or Medicine). We eliminated all prizes awarded for non-research reasons (e.g., teaching or history) or awarded to students. For the remainder, we searched the lists of winners against PubMed and if possible selected three articles by each prize winner. Awards that focus on biomedicine have deeper coverage in MEDLINE and so are better represented in our analysis than field-specific awards in areas peripheral to biomedicine (e.g., in ecology or entomology). We thus devoted more time to completing our coverage of articles associated with major or broad awards—which *should* be well represented in MEDLINE. These major and broad awards are thus more complete (i.e., more prize winners have associated articles) than narrower awards or those that only partially overlap the biomedical focus of PubMed.

We expanded the initial set of articles by finding the name of the prize winner in the MEDLINE entry for each of 1-3 "seed" articles. We then mapped that name using the Author-ity tool (36) to a cluster of partially disambiguated author names. We searched MEDLINE for the names in each author cluster. All retrieved publications written up to 30 years before the date of the award were associated with that prize winner. This process could introduce false positives for common names if we or our research assistants mis-assigned any seed articles, thereby associating a prize winner with an article written by another scientist (the Author-ity author clustering is quite conservative and is unlikely to contribute to this error). These false positives correspond to "typical" rather than award-winning scientists, however, and will likely behave like the rest of the typical population, causing us to underestimate the size of the prize effect. Prize winners with non-English characters in their names are also underrepresented; e.g., the Swedish Nobel Prize winner Sune Bergström is not well mapped to an Author-ity cluster and so some of his articles are not represented in our figures. Such false negatives also cause us to underestimate the enrichment of risky strategies in the strategy distribution of prize winners. Thus our results should be viewed as a conservative estimate of the difference between prize winners and the pool of all scientists.

The "Elite, general" prizes in Figure 4B were selected as follows. We included the two Nobel Prizes; prizes explicitly mentioned as being "precursors" to a Nobel; the top general awards given by professional societies in the US and the UK (the American Chemical Society, the National Academy of Science, and the Royal Society of Chemistry); and the Wolf Prizes, which have similar stature and prestige. The resulting list of 12 prizes is: Nobel Prize in Physiology or Medicine; Nobel Prize in Chemistry; Louisa Gross Horwitz Prize; Lasker-DeBakey Clinical Medical Research Award; Albert Lasker Award for Basic Medical Research; Gairdner Foundation International Award; NAS Award in Chemical Sciences; Priestley Medal; Corday-Morgan Medal; Grand Prix Charles Leopold Mayer; Wolf Prize in Medicine; Wolf Prize in Chemistry.

*See supporting information for additional technical details and discussion:*
http://dl.dropbox.com/u/9831990/Supporting_Information_Tradition_Innovation.pdf

**Acknowledgements**

We are grateful for helpful comments from Carl Bergstrom, Erica Cartmill and David Blei. We thank: Martin Rosvall and Daniel Edler for generous assistance with use of their MapEquation software; Jeff Alstott for help with his Python package powerlaw; research assistants Mahmoud Bahrani, Simo Huang, David Kates, and Val Michelman for help compiling data on prize winners in biomedicine and chemistry; and ThomsonReuters for making their citation information available. This work was supported by NSF grant SBE 0915730.

**Figures**

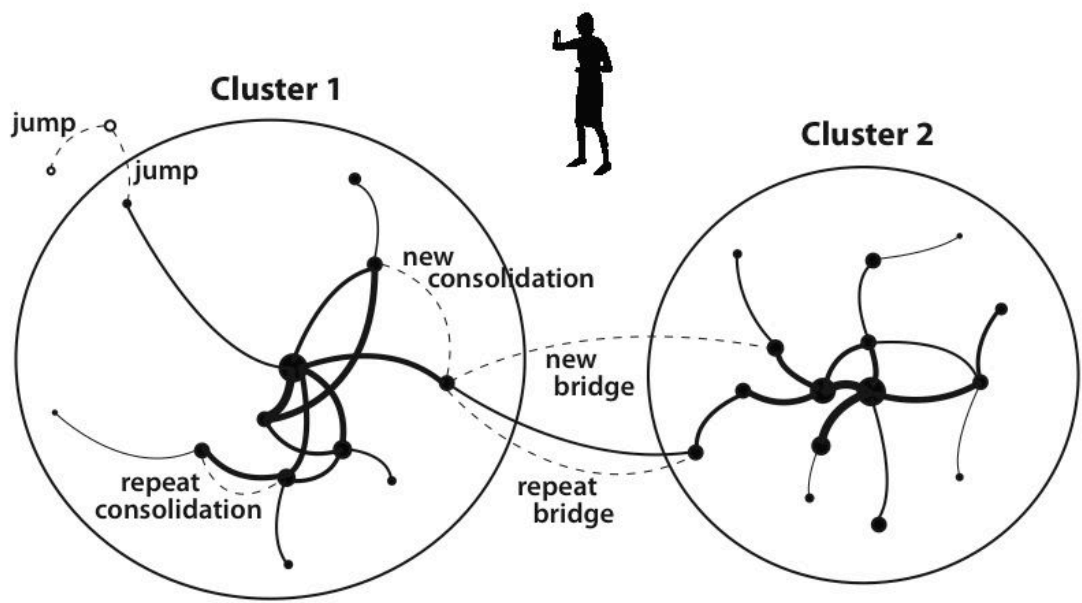

**Fig. 1.** Scientific strategies on a network. Nodes represent chemicals and links represent chemical relationships.

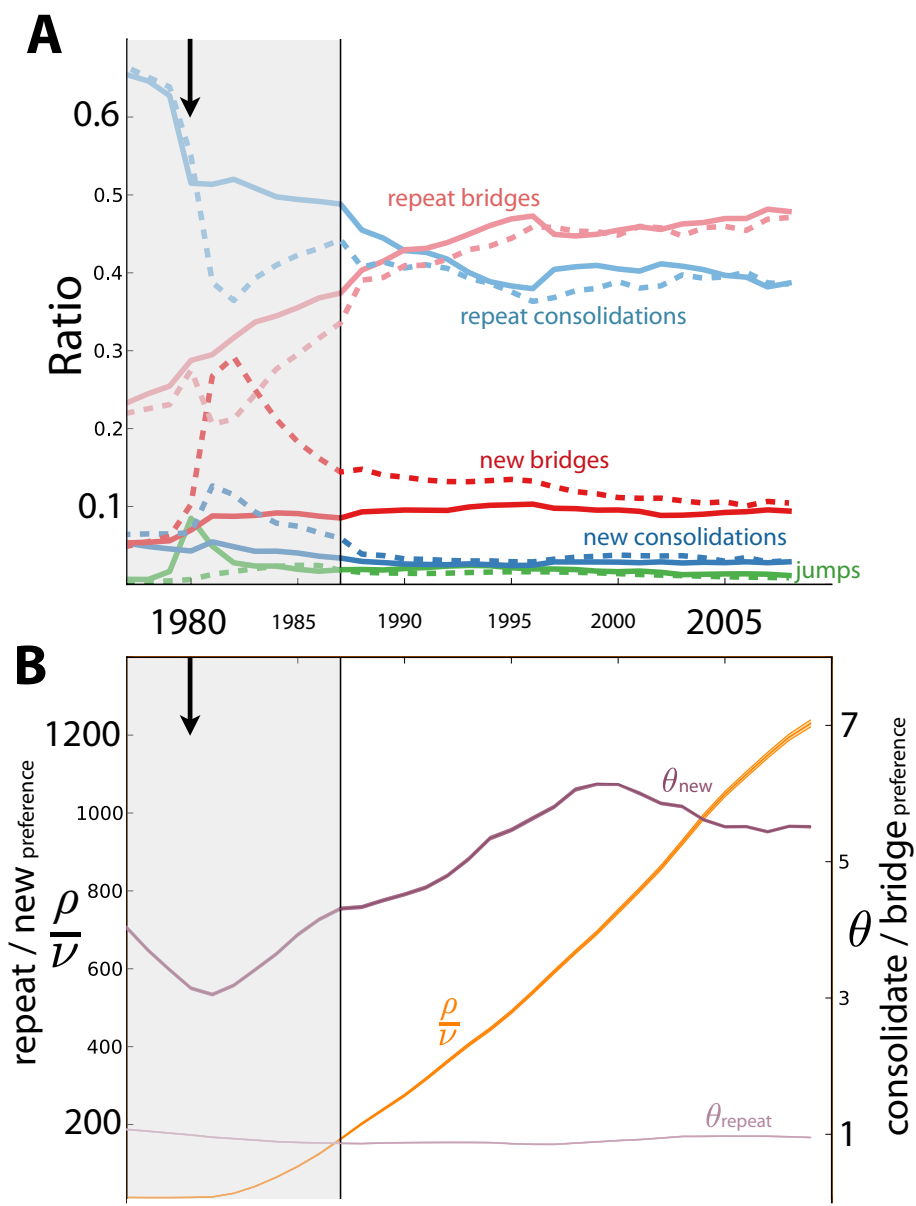

**Fig. 2.** Stable strategies and dynamical attention. **(A)** The empirical frequency of each strategy (solid line) with 95% confidence intervals smaller than the solid lines (*Materials and Methods*). Dotted lines show the predictions of our generative model. In 1980 (black arrow) chemical annotation is introduced in MEDLINE; see *SI*. This distorts parameter estimates until 1987 (parameters for year $t$ are inferred from the six previous years). **(B, left axis)** The ratio (orange) indicates an increasing concentration on established knowledge relative to new possibilities. **(B, right axis)** The preference for new consolidations over new bridges (plum) grows and stabilizes. This interest does not carry over to repeat work (lavender). 95% Bayesian credible intervals, shown in lighter colors, are very small (*Materials and Methods*).

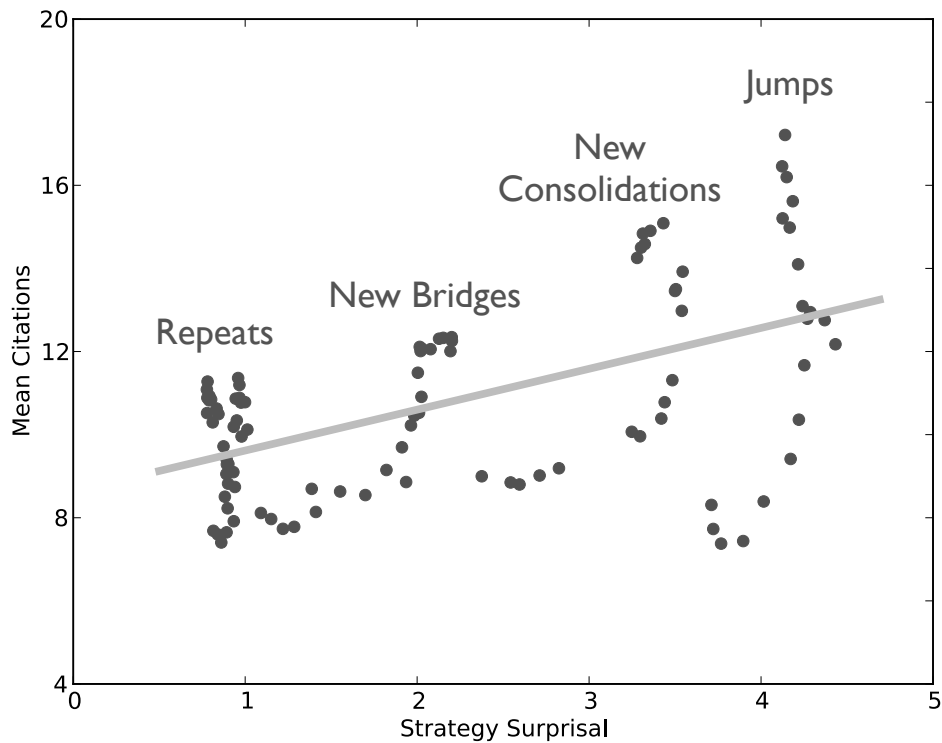

**Fig. 3.** Strategies and citation impact. Low-frequency strategies are correlated with higher mean citations (29% of variation explained). The surprisal of a strategy in year *t* (horizontal axis) plotted against mean citations received by papers published in year *t* containing at least one instance of that strategy. Much of the remaining variation is explained by adding year of observation to the model, as citations tend to increase with time (76% of variation explained by the combined model; the surprisal coefficient remains significant).

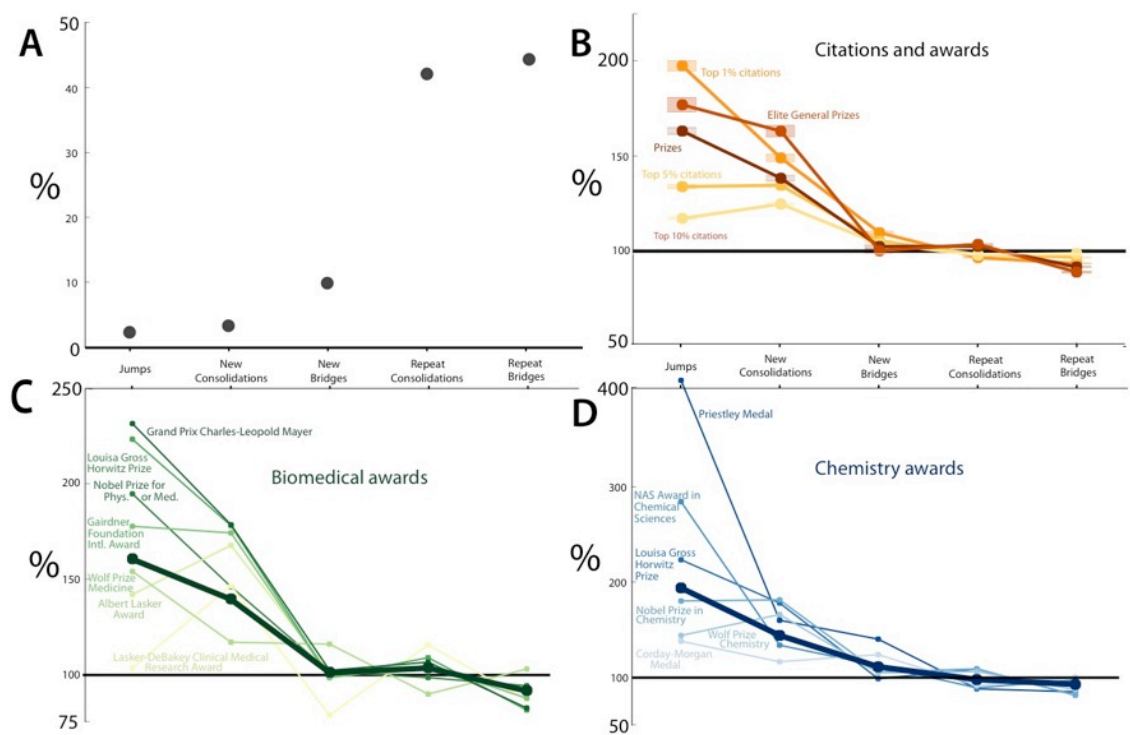

**Fig. 4.** Strategies shift for high impact scientists. **(A)** Pooling all papers published from 1983-2002, we find a distribution of strategies similar to that observed in any year (Fig. 2A). We judge pools of high impact articles and authors against this baseline (see Table S2). **(B)** Highly cited articles or those published by prize winners in the 30 years before receipt of a prize show an enhancement of jumps and consolidations. We assessed this for 137 prizes offered for research contributions in biology, medicine, and chemistry, representing 7594 awardees and 241,176 articles. The vertical axis measures the observed strategy frequency in each pool as a percentage of the baseline value. We emphasize that while the increase in frequency is small, relative to the baseline it is substantial. 95% confidence intervals are indicated by the boxes outside each data point. **(C)** and **(D)** As in **(B)** but for biological and medical prizes, and chemical prizes, respectively. The heavy line shows the aggregate behavior of all prizes; confidence intervals are smaller than the symbols (*Materials and Methods*). We also show the most prestigious, general prizes in these fields; confidence intervals not shown.